     \documentstyle[proceedings,psfig]{./crckapb} 
\begin{opening}

   \title{Towards the Modelling\protect\\
          of the Second Solar Spectrum}

   \author{Javier Trujillo Bueno}

   \institute{Instituto de Astrof\'\i sica de Canarias\protect\\
    E-38200, La Laguna,  
   Tenerife, Spain}

\end{opening}

\runningtitle{Modelling of the Second Solar Spectrum}

\begin{document}

% The \begin{document} command comes after the \end{opening}
% command.

\begin{abstract}\footnotesize\ 
This paper\footnote{Published in 1999 in the book {\it Solar Polarization}, edited by K.N. Nagendra \& J.O. Stenflo. Kluwer Academic Publishers, 1999. (Astrophysics and Space Science Library ; Vol. 243), p. 73-96} 
addresses the modelling issue of the second solar spectrum.
This is the name given to the linearly polarized spectrum
which can be observed close to the solar limb using spectro-polarimeters 
of high polarimetric sensitivity (Stenflo and Keller, 1997).
The second solar spectrum is due to scattering processes and offers a rich
diagnostic potential for exploring solar magnetic fields via the
Hanle effect. However, it is full of mysterious spectral features 
that cannot be understood with simplified polarization transfer theories,
thus suggesting that the underlying scattering physics is more complex than 
previously thought. In this paper we argue that understanding
the second solar spectrum requires the consideration of
scattering processes in {\it multilevel} atomic models, taking fully into account the transfer of {\it atomic polarization} among {\it all} the
levels involved. To give support to this statement, we begin 
by pointing out the 
drastically different predictions, given by
the standard resonance line polarization theory, with respect to
the emergent polarization in three
{\it different} line transitions. This standard theory neglects the
atomic polarization of the {\it lower} level of the
line transition under consideration, i.e. it assumes that
there are no population imbalances among the lower-level sublevels.
The density matrix polarization
transfer theory is then applied to
formulate the scattering polarization
problem taking properly into account atomic polarization in both the upper
and the lower line levels. The consideration of 
lower-level atomic polarization
leads to coupled {\it non-linear} and {\it non-local}
sets of equations, even for the two-level model atom
case considered in this paper. The {\it unknowns} of these
equations are the irreducible tensor components of the
atomic density matrix whose {\it self-consistent} values
have first to be obtained to be able to calculate the emergent Stokes profiles.
To solve this non-LTE problem 
of the $2^{nd}$ kind we present some iterative methods that 
are very suitable for developing a general
multilevel scattering polarization code. With these
numerical methods some model calculations are performed in order 
to demonstrate that the inclusion of lower-level atomic polarization leads
to similar emergent linear polarization signals in such
three {\it different} line transitions, as some observations 
show. After pointing out that the
``Na solar paradox'' (Landi Degl'Innocenti, 1998) might
admit, in principle, a {\it multilevel} solution,
the paper ends establishing
a new solar paradox: ``the Mg solar paradox'', for which no {\it multilevel}
solution seems to be possible. This new result demonstrates that
there indeed exists {\it ground} and {\it metastable}-level 
atomic polarization in the solar chromosphere and it suggests
that the solution to these ``solar paradoxes'' is to be found by 
carefully revising our current ideas about the chromospheric magnetic field.

%\keywords{\footnotesize
%polarization, scattering, magnetic fields, 
%methods: numerical, radiative transfer, sun: chromosphere, stars: %atmospheres}

\end{abstract}

\section{Introduction}

The ``second solar spectrum'' is a term adopted by
Stenflo and Keller (1997) to refer to the remarkable observational
discovery that the whole solar spectrum is linearly polarized, 
when observations are made close to the solar limb using novel
spectro-polarimeters that allow the detection of
very low amplitude polarization signals
($\sim10^{-5}$ in the degree of linear polarization).  
This term is certainly adequate because, as pointed out by these authors,
the linearly polarized solar spectrum has a structural richness
that often exceeds that of the ordinary intensity spectrum. It is
indeed as if the Sun has presented us with an entirely new spectrum to explore.
In fact, the second solar spectrum contains a wealth of ``inexplicable''
spectral features which are the signature of physical processes
that are presently challenging physicists working in the field.

This paper deals with the modelling issue of the second
solar spectrum. This modelling requires the solution of 
a formidable numerical problem that is considered 
as one of the most challenging tasks of 
solar and stellar polarimetry. It consists in
calculating, for {\it multilevel} atomic models,
the excitation and ionization states of chemical species of given abundance
that are {\it consistent} with the polarization properties
of the radiation field produced by such species in any medium of given
temperature, density, macroscopic velocity and magnetic field vector.
Once this self-consistent atomic excitation is known along the line of sight,
it is straightforward to solve the transfer equations for the Stokes
parameters in order to calculate the emergent polarization profiles
that are to be compared with spectro-polarimetric observations.

It is indeed a very complex problem because, in the polarization transfer
case, one has to take into account that
each level of total angular momentum value X has associated with it
(2X+1) sublevels, with ${\vec {\rm X}}={\vec {\rm J}}=
{\vec {\rm L}}+{\vec {\rm S}}$ if fine 
structure due to the spin-orbit LS coupling is assumed, or
with ${\vec {\rm X}}={\vec {\rm F}}={\vec {\rm J}}+{\vec {\rm I}}$ 
if hyperfine structure 
due to the nuclear angular momentum $\vec {\rm I}$ is taken into account.
The populations of these sublevels are sensitive
to the polarization and anisotropy state of the radiation field at each point within the medium. Moreover, quantum interferences
(or coherences) among the sublevels themselves may also appear,
coherences that depend on the energy separation between the levels
and on their splitting. These coherences  
must also be properly quantified to fully specify the excitation state. 
An additional complication stems from the fact that,
in the polarized case, instead of the 
standard radiative transfer (RT)
equation for the specific intensity, one has to
solve, in general, a vectorial transfer equation for the 
four Stokes parameters.

Obviously, accounting for this complexity requires working within the framework of a robust theory for the generation and transfer of polarized
radiation. I believe that the most rigorous (and suitable)
theoretical framework to work with is that
developed by Landi Degl'Innocenti (1983), which is based
on the irreducible tensor components ($\rho^K_Q$) of the
atomic density matrix (see also Bommier and Sahal-Br\'echot, 1978).
According to this formalism, to each level of angular momentum
value X, there correspond $\rm (2X+1)^{2}$ density-matrix elements.
These $\rho^K_Q$-elements contain information about the populations of the 
atomic sublevels and about the coherences among them. For instance,
for a level with total angular momentum
J=1 we have that $\rho_0^0=(N_1+N_0+N_{-1})/{\sqrt{3}}$,
$\rho^1_0=(N_1\,-\,N_{-1})/{\sqrt{2}}$ and
$\rho_0^2=(N_1-2N_0+N_{-1})/{\sqrt{6}}$, where $N_{i}$ 
(with $i$=1, 0 and -1) are the populations of the three magnetic sublevels.
Thus, $\rho_0^0$ gives the total population of the level, while
$\rho^1_0$ (the {\it orientation} coefficient) and $\rho^2_0$ 
(the {\it alignment} coefficient) inform us about the {\it population differences} among the sublevels. Finally, $\rho^K_Q$-terms with $Q\neq0$
account for the coherences between Zeeman sublevels
whose magnetic quantum numbers differ by $Q$.

One of the reasons that explain why 
the density matrix formalism is so suitable
for dealing with the generation and transfer of polarized radiation
is that, through an emission process, polarization in spectral lines
can originate locally either by the splitting
of the atomic levels (splitting that can in turn be due either to the Zeeman
or the Stark effect) or by the presence of
{\it population differences} and/or {\it coherences} among the sublevels.
Thus, the $\rho^K_Q$ elements
whose self-consistent values are sought at each spatial grid-point,
indeed provide the most suitable way of quantifying, at the atomic level, 
the information that we need to be able to calculate all the
``sources'' and ``sinks'' of polarization within the medium under consideration. The main criticism of this QED theory is that it is based on the approximation
of complete frequency redistribution (CRD), i.e. on an assumption that is
not adequate for modelling the polarization of several diagnostically
important spectral lines. Fortunately, some very recent work
has successfully started to incorporate the effects of partial
redistribution (PRD) into the framework of the density matrix formalism
(see Landi Degl'Innocenti {\it et al.}, 1997; Bommier, 1997 a,b).
These recent efforts to generalize the density matrix theory to PRD
are truly important and should be continued because the CRD theory
cannot be applied when coherences between 
non-degenerate levels are present unless the spectrum of the radiation
is flat across a frequency interval $\Delta\nu$ centred on the line frequency and larger than the frequency separation between the two levels connected by the coherence (see Landi Degl'Innocenti {\it et al.}, 1997).

The problem of finding the self-consistent values
of the irreducible tensor components of the atomic density matrix
has been called ``the non-LTE problem of the $2^{nd}$ kind''
(see Landi Degl'Innocenti, 1987). It requires
solving jointly the statistical equilibrium (SE) equations
for the density matrix elements associated with each level of the assumed 
atomic model and the Stokes-vector transfer equations
for all the radiative transitions involved. This terminology also seems 
appropriate because, as I shall try to argue below, 
a better understanding
of many of the ``inexplicable'' spectral features of the 
second solar spectrum
can only be achieved by carefully formulating and solving {\it multilevel}
non-LTE problems of the $2^{nd}$ kind, taking fully into account
the possibility of {\it atomic polarization} at {\it all} the levels
of the chosen multilevel atomic model and including
the depolarizing role of elastic collisions and magnetic fields.

The word ``inexplicable'' in the preceding paragraphs 
refers to the impossibility of explaining some of the spectral features 
of the second solar spectrum by means of theories
based on the approximation of neglecting atomic polarization
in the {\it lower} level of the line transition under consideration,
i.e. theories based on the assumption that there are no 
significant differences in the populations of the lower-level sublevels or coherences among them. One example of a mysterious feature
of the second solar spectrum that has triggered some recent theoretical work 
(see Trujillo Bueno and Landi Degl'Innocenti, 1997;
Landi Degl'Innocenti, 1998) is the linear polarization
pattern observed around the Na I ${\rm D}_2$ and ${\rm D}_1$ lines. 
In fact, in a recent 
letter in {\it Nature} that demonstrates the robustness of the density matrix
polarization transfer theory including partial 
frequency redistribution effects, 
Landi Degl'Innocenti (1998) concluded 
that the observed Na ${\rm D}_2$ and ${\rm D}_1$
linear polarization pattern can be explained by
assuming the presence of an amount of {\it ground}-level atomic
polarization as important as (and indeed slightly larger than!) 
that of the upper level. However, because very small 
{\it non-vertical} magnetic fields (and/or elastic collisions)
destroy the ground-level atomic polarization of Na, Landi Degl'Innocenti (1998) 
was forced to rule out in the solar chromosphere
both elastic collisions and the existence of
turbulent magnetic fields and of horizontal, canopy-like fields stronger than $\sim0.01$ gauss. This has led to 
an exciting apparent paradox in solar physics because
there are observational evidence for both
turbulent fields of the order of 10 gauss and canopy-like horizontal fields
(see Jones, 1984; Solanki and Steiner, 1990; Faurobert-Scholl, 1992;
Bianda {\it et. al}, 1998; Stenflo {\it et. al}, 1998).

The argument in favour of the simplifying approximation 
of neglecting lower-level polarization
is that the lower level of a line transition is generally long-lived, 
and that it must thus have plenty of time to be depolarized by elastic collisions and/or weak magnetic fields (Stenflo, 1994; 1997). However, 
this is expected to be the case concerning only the {\it ground} 
and {\it metastable} levels of atomic systems.
It is indeed a major simplifying approximation because it implies that
the scattered radiation can be expressed linearly in terms of the incident
radiation if stimulated emission processes are also neglected (see Landi Degl'Innocenti, 1984). In other words, the approximation 
of neglecting lower-level atomic polarization allows the study
of scattering line polarization problems 
in terms of phase matrices that are decoupled from the SE equations. 
However, as we shall show below, the consideration
of lower-level atomic polarization leads to a coupled system of
{\it non-linear} equations, even for the simplest 
case of a two-level model atom.

My motivation for writing this paper is two-fold.
Firstly, I would like to provide more arguments concerning
the idea (already put forward in the paper by Trujillo
Bueno and Landi Degl'Innocenti, 1997)
that lower-level atomic polarization is an essential physical
ingredient for understanding the second solar spectrum.
Secondly, I aim to demonstrate that,
contrary to some general beliefs,
the density-matrix polarization transfer theory
does have a suitable form for practical applications.

To these ends, I will consider here three types of line transitions
in the solar atmosphere: ($a$) lines with ${\rm J}_{l}=0$ and
${\rm J}_u=1$, ($b$) lines with ${\rm J}_{l}=1$ and
${\rm J}_u=0$, and ($c$) lines with ${\rm J}_{l}=1$ and ${\rm J}_u=1$.
Section 2 summarizes the predictions of the standard theory,
which neglects lower-level atomic polarization.
Section 3 is dedicated to outlining the formulation of the
scattering line polarization problem taking into account atomic
polarization in both the upper and the lower levels of such line transitions.
Here I will show the self-consistent solution for the density matrix elements
and the corresponding emergent fractional linear 
polarization that results from this
more correct treatment. Section 4 discusses
model calculations including the depolarizing role of elastic collisions.

As we shall see, my two-level atom  
scattering line polarization calculations suggest that,
if we aim at understanding the second solar spectrum,
we need to consider scattering processes
in {\it multilevel} atomic models, taking fully into account
the transfer of atomic polarization among all the atomic levels involved.
Section 5 argues that, in principle, the above-mentioned
``Na solar paradox'' might admit a {\it multilevel} solution.
However, Section 6 shows 
that the observed fractional linear polarization
in the Mg $b$-lines can only be explained by invoking the presence
of atomic polarization in the lower {\it metastable} levels
of the Mg $b_1$ and $b_2$ lines, thus establishing
a new paradox: the ``Mg solar paradox''. Finally, Section 7
gives some concluding remarks after pointing out
that there is no multilevel solution to this ``Mg solar paradox''.
The Appendix is dedicated
to a brief description of some iterative methods 
for the solution of non-LTE
problems of the $2^{nd}$ kind, which I consider
as ``the road to be taken'' for the development of a general multilevel
scattering line polarization code. 
 
\section{Predictions of the Standard Theory}

This and the following section deal with
the scattering line polarization problem 
assuming a one-dimensional (1D), plane-parallel, non-magnetic,
static solar atmospheric model and a two-level 
model atom neglecting stimulated emission processes.
The only difference is that the results presented in this section
are obtained with the standard theory, which neglects
{\it atomic polarization} in the lower level,
while those shown in the next section are the result
of the application of the full theory that takes into account atomic 
polarization at both levels.

A detailed formulation of the {\it standard} resonance
line polarization problem, together with some 
numerical results obtained applying
novel iterative schemes, can be found in the paper by
Trujillo Bueno and Manso Sainz (1999). In that paper 
it is clarified that in order to specify 
the excitation state of two-level atoms without
ground-level atomic polarization
it is enough to consider two density-matrix elements
at each spatial grid-point: $\rho_0^0(u)$ and $\rho_0^2(u)$
(see also Landi Degl'Innocenti {\it et. al.}, 1990).
The first one, i.e. $\rho^0_0(u)$,
measures the overall population of the {\it upper} level,
while $\rho_0^2(u)$ is the so-called {\it alignment} coefficient,
which quantifies the degree of imbalance 
in the populations of the {\it upper}-level sublevels. 

In the stellar atmospheric environment
the main physical mechanism that leads to 
population differences among the sublevels of the atomic levels
is the anisotropic illumination
of the atoms. This is easy to understand by considering the academic
case of a unidirectional unpolarized light beam 
that illuminates a gas of two-level
atoms with ${\rm J}_l=0$ and ${\rm J}_u=1$
and that is propagating along the
direction chosen as the quantization axis.
Since these atoms must absorb $\pm1$ units of angular momentum from
the light beam, only the transitions corresponding
to $\Delta{M}=\pm 1$ are effective, 
so that no transitions occur to the $M=0$ sublevel of the upper level.
Thus, in the absence of any relaxation mechanisms,
the upper-level sublevels with $M=1$ and $M=-1$ would be more populated than
the $M=0$ sublevel and the alignment coefficient 
$\rho^2_0(u)=(N_1\,-2N_0\,+N_{-1})/{\sqrt{6}}$ would
have a {\it positive} value. 
Clearly, the amount of this atomic alignment in a
stellar atmospheric environment is significantly smaller than
in such an academic case due to the relaxation mechanisms
present in a stellar atmosphere 
(e.g. depolarizing collisions and magnetic fields)
and to the much weaker degree of the radiation field's anisotropy.

As mentioned above,
in order to calculate the fractional linear polarization
emerging from a given solar atmospheric model we have first to
find the {\it self-consistent} values of $\rho^0_0(u)$ and $\rho^2_0(u)$
by solving the SE and the RT equations (see Trujillo Bueno and Manso
Sainz, 1999). Figure 1$a$ shows the variation with the line integrated
optical depth of the self-consistent values of $\rho^2_0(u)/\rho^0_0(u)$
for the three types of line transitions mentioned in the introduction,
calculated for the case of {\it zero}
depolarizing rate due to elastic collisions.

\begin{figure}[htb]
\centerline{\psfig{file=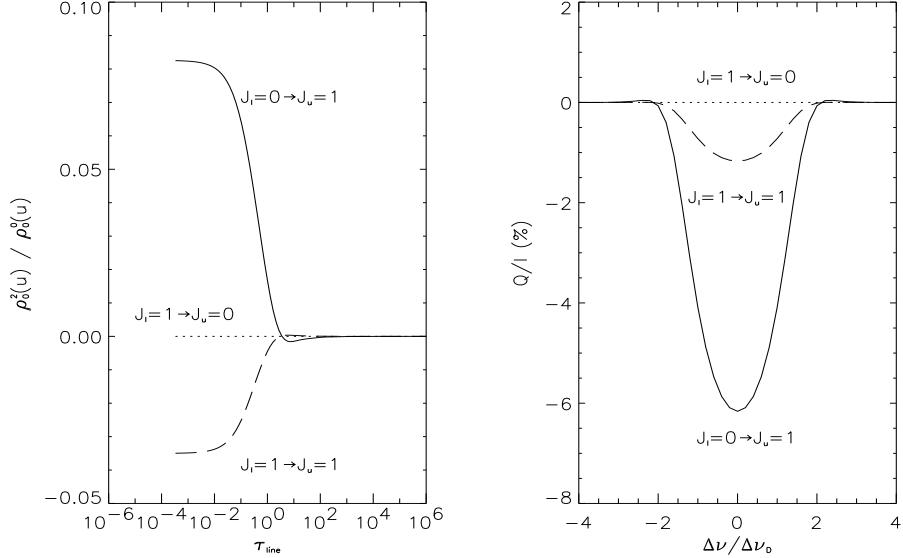,width=12.5cm,height=8cm}}
\label{fig1}
\caption[fig1]{Predictions of the standard theory
(which neglects lower-level polarization)
for the zero depolarizing rate case in
three different line transitions. The left panel (Fig. 1$a$) shows
the variation
with line integrated optical depth of $\rho^2_0(u)/\rho^0_0(u)$.
The right panel (Fig. 1$b$) shows 
the emergent fractional linear polarization
at $\mu=0.1$, with
positive Q defined along the radial direction
through the observed point.
The input atmospheric model is an isothermal atmosphere with the
solar effective temperature. Inelastic collisions are taken
into account by assuming a collisional destruction probability
$\epsilon=10^{-4}$.}
\end{figure}

As seen in Fig. 1$a$, the largest values for the
upper-level alignment are obtained for line transitions with
${\rm J}_l=0$ and ${\rm J}_u=1$, i.e. for triplet lines. Alignment values
$\sim\,8\,\%$ are found in the surface layers, while
$\rho^2_0(u)$ becomes negligible for optical depths $\tau_{line}>1$,
because in these high-opacity regions the radiation field
approaches isotropy. The alignment of the upper level is found to be zero
for lines with ${\rm J}_l=1$ and ${\rm J}_u=0$,
as expected from the fact that levels with J=0 have no sublevel
structure. For lines with ${\rm J}_l={\rm J}_u=1$
we have $\rho^2_0(u)/\rho^0_0(u)\,\sim\,-3\%$, i.e. alignment
absolute values that are {\it smaller} than for the case of triplet lines.

Figure 1$b$ shows the corresponding emergent fractional linear
polarization (Q/I) for simulated observations 
at $\mu={\rm cos}\,{\theta}=0.1$, with $\theta$ 
being the angle between the line of sight and the normal 
to the stellar surface. 
In order to understand these results we point out that, for this
{\it standard} resonance line polarization case where population differences
among the lower-level sublevels are neglected, one has the following RT equations (see, e.g., Trujillo Bueno and Manso Sainz, 1999):

\begin{equation}
{\rm {{d}\over{ds}}\,{I}=
{{\epsilon_{I}}}-
{{\eta_{I}}}\,\,{I}},
\end{equation}   

\begin{equation}   
{\rm {{d}\over{ds}}\,{Q}=
{{\epsilon_{Q}}}-{{\eta_{I}}}\,\,{Q}},
\end{equation}
where $s$ is the geometrical distance
along the ray and where the {\it line} contributions to the 
I and Q components of the emission vector are:

\begin{equation}
{\epsilon_{\rm I}}^{line}\,=\,\,(h\nu/4\pi)\,A_{ul}{\cal N}
\,\sqrt{2{\rm J}_u+1}\,\,{\phi}_{x}\,[{{\rho}^{0}_{0}(u)}\,
+\,{{{\omega}}
\over{2\sqrt{2}}}\,(3\,\mu^2-1)\,\,{{\rho}^{2}_{0}(u)}],
\end{equation}

\begin{equation}
{\epsilon_{\rm Q}}^{line}\,=\,\,(h\nu/4\pi)\,A_{ul}{\cal N}
\,\sqrt{2{\rm J}_u+1}\,\,{\phi}_{x}\,{{3\,\,{\omega}}
\over{2\sqrt{2}}}\,(\mu^2-1)\,\,{{\rho}^{2}_{0}(u)},
\end{equation} 
with $\cal N$ the total number of atoms per unit volume,
$\nu$ the line frequency,
${A}_{ul}$ the Einstein spontaneous emission coefficient,
${\phi}_{x}$ the normalized 
line profile (where $x$ is the frequency
measured from the line centre in units of the Doppler width), and 
$\omega$ is equal to the $w^{(2)}_{{\rm J}_u,{\rm J}_l}$
coefficient introduced by Landi Degl'Innocenti (1984),
whose values for transitions 
${\rm J}_l\,\rightarrow\,{\rm J}_u$ are
(see Table 1 in Landi Degl'Innocenti, 1984):

$${\omega}\,=\,1\,\,\,\,{\rm for}\,\,\,0\,\rightarrow\,1$$

$${\omega}\,=\,0\,\,\,\,{\rm for}\,\,\,1\,\rightarrow\,0$$

$${\omega}\,=\,-1/2\,\,\,\,{\rm for}\,\,\,1\,\rightarrow\,1$$

The above transfer equations show
that, for this standard resonance line polarization case, 
the absorption matrix is {\it diagonal},
i.e. ${\bf K}={\eta_{\rm I}}\,{\bf 1}$, where the line contribution
to the I-component of the absorption matrix is given by

\begin{equation}
{\eta_{\rm I}}^{line}\,=\,\,(h\nu/4\pi)\,B_{lu}{\cal N}
\,\sqrt{2{\rm J}_l+1}\,\,{\phi}_{x}\,{{\rho}^{0}_{0}(l)},
\end{equation} 
where ${B}_{lu}$ is the Einstein coefficient
for the absorption process and
$\rho^0_0(l)$ is simply proportional to the total population
of the {\it lower} level.

Therefore, from Eqs. (2) and (4),
one sees that the emissivity in Q at each point within the atmosphere
is directly proportional to the alignment coefficient of the {\it upper}
level, weighted by a number ($\omega$)
that depends on the total angular momentum
of the lower and upper levels. The conclusion is that the standard
theory predicts the largest fractional linear polarization for triplet
lines ($\sim\,6\%$ at the line centre
at $\mu=0.1$), {\it zero} linear polarization for lines
with ${\rm J}_l=1$ and ${\rm J}_u=0$, while for lines
with ${\rm J}_l={\rm J}_u=1$  we have Q/I$\,\sim 1\%$
at $\mu=0.1$.

\section{The Effect of Lower-level Depopulation Pumping}

Consider again the above-mentioned academic case of an unpolarized
light beam incident on a gas of two-level atoms, but now with
${\rm J}_l=1$ and ${\rm J}_u=0$. In this case no transitions can occur
out of the $M=0$ sublevel of the {\it lower} level, since only transitions
corresponding to $\Delta {M}=\pm1$ are effective. On the other
hand, the spontaneous de-excitation from the upper level
populates with equal probability the three sublevels
($M=-1,0,+1$) of the {\it lower} level. In the absence of 
any relaxation mechanisms, the final result of this optical-pumping
cycle is that all atoms will eventually be pumped into the $M=0$ sublevel
of the {\it lower} level, and the medium will become transparent.

In the quantum optics literature this is known as {\it depopulation
pumping} (Happer, 1972; see also Landolfi and Landi Degl'Innocenti, 1986). 
As we have seen, depopulation pumping occurs when certain
lower-level sublevels absorb light more strongly than others.
As a result, an excess population tends to build up in the weakly
absorbing sublevels.
In the above two-level atom example depopulation pumping 
leads to a {\it negative} lower-level alignment
coefficient $\rho^2_0(l)\,=\,-2\,N_l/{\sqrt{6}}$,
with $N_l$ the total population of the lower level.
The physical mechanism that is responsible
of these population differences between the {\it lower}-level
subleves is again the {\it anisotropic} illumination of the atoms.
Thus, if the anisotropy of the solar radiation field is capable
of inducing significant population differences among the upper-level
sublevels, why should it not be capable of producing similar
{\it lower}-level alignment coefficient values? In fact, it does,
as demonstrated by Trujillo Bueno and Landi Degl'Innocenti (1997),
who formulated this scattering line polarization
problem and solved the ensuing non-local and 
non-linear set of
equations. For instance, for a line transition with ${\rm J}_l=1$ and ${\rm J}_u=0$ in a two-level model atom the rate equation
that governs the temporal evolution of $\rho^2_0(l)$ is

\begin{equation}
{{d}\over{d\,t}}\,{{\rho}^2_0(l)}\,=\,
-B_{lu}\,{{{\bar{J}}^{2}_{0}}}
{{{\rho}^{0}_{0}}(l)}+
{{B_{lu}}\over{\sqrt{2}}}{{{\bar{J}}^{2}_{0}}}
{{{\rho}^{2}_{0}}(l)}-
B_{lu}{{{\bar{J}}^{0}_{0}}}
{{{\rho}^{2}_{0}}(l)}-(C_{lu}+D_{l})
{{{\rho}^{2}_{0}}(l)}=0,
\end{equation}
where $C_{lu}$ is the 
upward inelastic collisional rate and $D_l$ the lower-level
depolarizing rate due to elastic collisions. In this equation
${{\bar{J}}^{0}_{0}}$ is given by a frequency and angular
average of the Stokes-I parameter weighted
by the line absorption profile, while 
${{\bar{J}}^{2}_{0}}$ is given by a frequency and angular
integral of the Stokes I and Q parameters, 
additionally weighted by some angle-dependent
factors. These quantities 
(${{\bar{J}}^{0}_{0}}$ and ${{\bar{J}}^{2}_{0}}$)
are two of the spherical tensor
components of the radiation field 
(see Landi Degl'Innocenti, 1983) and they
quantify, respectively, the average intensity of the radiation field 
and mainly the degree of its anisotropy.

All the rates in Eq. (6) are {\it relaxation} rates. 
We point out that the most important radiative rates here are the
first and the third, which are due, respectively, to the {\it anisotropic}
(${\bar{J}}^{2}_{0}$) and to the 
{\it isotropic} (${\bar{J}}^{0}_{0}$) components
of the radiation field tensor, acting on the unpolarized
and polarized components of the lower-level density matrix,
respectively. Neglecting
in Eq. (6) the collisional rates and taking into account the
weakly anisotropic nature of the solar radiation field (i.e. that
${\bar{J}}^{2}_{0}/{\bar{J}}^{0}_{0}\,\ll\,1$) one finds
that $\rho^2_0/\rho^0_0\,\sim\,-{\bar{J}}^{2}_{0}/{\bar{J}}^{0}_{0}$.
Figure 2$a$ shows the self-consistent solution for $\rho^2_0(l)/\rho^0_0(l)$,
for the zero depolarizing rate case, but taking into account inelastic collisions.

The question now is whether the presence of population differences
in the lower level can by itself lead to local sources of linear
polarization. The answer is affirmative. To understand this we have
to write down the transfer equations for the Stokes I and Q parameters
that result from the application of the density-matrix theory
taking into account the possibility of atomic polarization
in both levels (Trujillo Bueno and Landi Degl'Innocenti, 1997):

\begin{equation}
{\rm {{d}\over{ds}}\,{I}=
{{\epsilon_{I}}}-
{{\eta_{I}}}\,\,{I}-{{\eta_{Q}}}\,\,{Q}},
\end{equation}   

\begin{equation}   
{\rm {{d}\over{ds}}\,{Q}=
{{\epsilon_{Q}}}-{{\eta_{Q}}}\,\,{I}-{{\eta_{I}}}\,\,{Q}},
\end{equation}
where the {\it line} contribution to
$\eta_{\rm I}$ and the full
Q-component of the absorption matrix (which is {\it not}
diagonal now !) are given by

\begin{equation}
{\eta_{\rm I}}^{line}\,=\,\,(h\nu/4\pi)\,B_{lu}{\cal N}
\,\sqrt{2{\rm J}_l+1}\,\,{\phi}_{x}\,
[{{\rho}^{0}_{0}(l)}+{{\cal Z}
\over{2\sqrt{2}}}\,(3\mu^2-1)\,\,{{\rho}^{2}_{0}(l)}],
\end{equation}

\begin{equation}
{\eta_{\rm Q}}\,=\,\,(h\nu/4\pi)\,B_{lu}{\cal N}
\,\sqrt{2{\rm J}_l+1}\,\,{\phi}_{x}\,
{{3\,\,{\cal Z}}
\over{2\sqrt{2}}}\,(\mu^2-1)\,\,{{\rho}^{2}_{0}(l)},
\end{equation}
where $\cal Z$ should not be confused with the $\omega$-symbol
of Eqs. (3) and (4), since $\cal Z$ is equal to the symbol
$w^{(2)}_{{\rm J}_l,{\rm J}_u}$ introduced by Landi Degl'Innocenti (1984),
while $\omega=w^{(2)}_{{\rm J}_u,{\rm J}_l}$. The values of $\cal Z$
for transitions ${\rm J}_l\,\rightarrow\,{\rm J}_u$ are:

$${\cal Z}\,=\,0\,\,\,\,{\rm for}\,\,\,0\,\rightarrow\,1$$

$${\cal Z}\,=\,1\,\,\,\,{\rm for}\,\,\,1\,\rightarrow\,0$$

$${\cal Z}\,=\,-1/2\,\,\,\,{\rm for}\,\,\,1\,\rightarrow\,1$$

Equation (4) shows that $\epsilon_{\rm Q}^{line}\,=\,0$
for $1\,\rightarrow\,0$ transitions, as it must be, because
for a level with ${\rm J}=0$ the alignment is zero. Thus, 
for $1\,\rightarrow\,0$ transitions the term $-\eta_{\rm Q}\,{\rm I}$
in Eq. (8) is the only one that plays the role of the emissivity in Q at each
point in the atmosphere. In other words, Eq. (10) shows that the existence
of differences in the populations of the magnetic sublevels
of the {\it lower} level leads to non-zero values of
${\eta}_{\rm Q}$, i.e. it introduces {\it dichroism} in the
stellar atmosphere and a coupling of the intensity of the radiation
beam with the
Stokes Q-parameter that is due to the absorption process.
Note that the larger
the absolute value of the alignment coefficient $\rho^2_0(l)$,
the larger the expected Q/I signal (Trujillo Bueno and Landi Degl'Innocenti, 1997). Figure 2$b$ shows, for the {\it zero}
depolarizing rate case, the emergent fractional linear polarization
at $\mu=0.1$ that corresponds to the self-consistent solution
given in Fig. 2$a$. In conclusion, the prediction of the correct 
two-level atom theory for line transitions with 
${\rm J}_l=1$ and ${\rm J}_u=0$ is that the 
emergent fractional linear polarization
is {\it as important as that corresponding to triplet lines}.

\begin{figure}[htb]
\centerline{\psfig{file=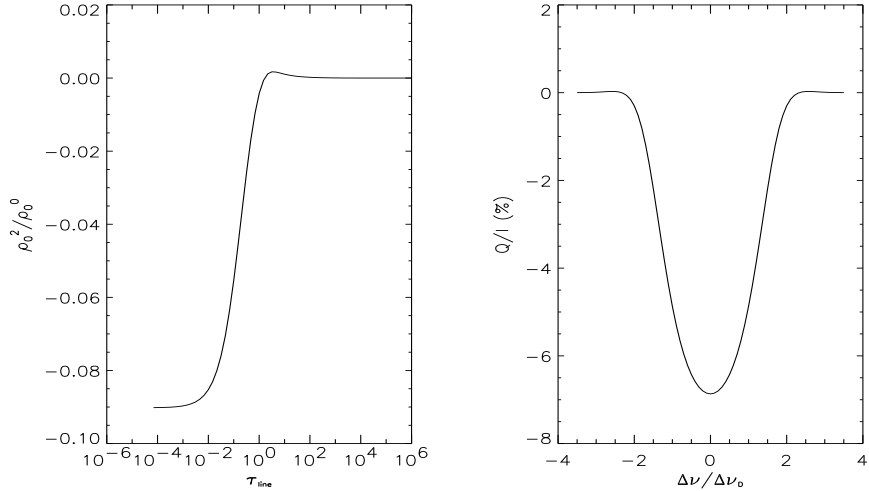,width=12cm,height=7cm}}
\label{fig1}
\caption[fig1]{Predictions of the full theory
for line transitions in a two-level model atom
with ${\rm J}_l=1$ and ${\rm J}_u=0$. The left panel (Fig. 2$a$) shows
the variation
with line integrated optical depth of $\rho^2_0(l)/\rho^0_0(l)$.
The right panel (Fig. 2$b$) shows 
the emergent fractional linear polarization
at $\mu=0.1$, which is to be compared with the dotted-line
of Fig. 1$b$.}
\end{figure}

Finally, we ask 
how important can be the fractional linear 
polarization in lines with ${\rm J}_l={\rm J}_u=1$ when
atomic alignment is taken into account in both 
the upper and the lower levels.
For two-level atoms with these angular momentum values
four quantities are needed
to fully specify their excitation state: $\rho^0_0(u)$,
$\rho^2_0(u)$, $\rho^0_0(l)$ and $\rho^2_0(l)$. These quantities
at all the spatial grid-points are the {\it unknowns} whose
self-consistent values are sought. To this end, one has
to derive the SE equations applying the density-matrix theory. 
Now, corresponding to each spatial grid-point,
one has four equations with four unknowns.

A detailed presentation of these {\it non-linear} equations will be published
elsewhere. For our purposes here we simply write down the rate
equation that governs the temporal evolution of $\rho^2_0(u)$:

\begin{eqnarray}
{{d}\over{d\,t}}\,{{\rho}^2_0(u)}\,=\,
-A_{ul}\,{{{\rho}^{2}_{0}}(u)}
-{B_{lu}\over2}{{{\bar{J}}^{2}_{0}}}
{{{\rho}^{0}_{0}}(l)}-
{{B_{lu}}\over{{2}}}{{{\bar{J}}^{0}_{0}}}
{{{\rho}^{2}_{0}}(l)}-
{B_{lu}\over{\sqrt{2}}}{{{\bar{J}}^{2}_{0}}}
{{{\rho}^{2}_{0}}(l)} \\
+C^{(2)}_{lu}{{{\rho}^{2}_{0}}(l)} 
-(C_{ul}+D_{u})
{{{\rho}^{2}_{0}}(u)} = 0.
\nonumber
\end{eqnarray}
In this equation the first and last terms are {\it relaxation} rates,
while the remaining ones are {\it transfer} rates. We point out that,
besides the usual rate ${\cal T}_2=-{B_{lu}}{{{\bar{J}}^{2}_{0}}}
{{{\rho}^{0}_{0}}(l)}/2$, which results from the {\it anisotropic}
component (${\bar{J}}^{2}_{0}$) of the radiation field,
we now have three {\it extra}
rates that are due to {\it transfer of atomic polarization
from the lower level to the upper level}. The most important 
of these three rates is
the transfer rate due to the {\it isotropic} component (${\bar J}^0_0$)
of the radiation field, i.e. 
${\cal T}_3=-{{B_{lu}}}{{{\bar{J}}^{0}_{0}}}{{{\rho}^{2}_{0}}(l)}/2$.
The rate ${\cal T}_5=C^{(2)}_{lu}{{{\rho}^{2}_{0}}(l)}$ is to be interpreted
as transfer of atomic alignment from the lower to the upper level by
{\it inelastic} collisions, where $C^{(2)}_{lu}$ is the $K=2$ multipole
component of the inelastic collisional rate. These three {\it extra} rates (but
mainly ${\cal T}_3$!) produce an additional contribution to the atomic
alignment of the upper level which, in turn, 
implies an {\it extra} contribution to the
linear polarization emitted by the atoms at each spatial point.
It is also important to point out that there is a further mechanism
that contributes to the emergent linear polarization, and that
it also arises from the presence of atomic alignment in the lower level.
This is due to the
term $-\eta_{\rm Q}{\rm I}$ in Eq. (8), with $\eta_{\rm Q}$ given
by Eq. (10), i.e. to the differential absorption due to the
unequally populated sublevels of the lower level. As discussed above,
this is the only mechanism that leads to linear polarization
in $1\,\rightarrow\,0$ transitions.

The transfer
equations for the Stokes I and Q parameters are given by Eqs.
(7) and (8), with the components of the emission vector and
of the absorption matrix given by Eqs. (3),(4),(9) and (10). (Note that,
for the present case of $1\,\rightarrow\,1$ transitions, the coefficients
$\omega$ and $\cal Z$ are such that ${\omega}={\cal Z}=-0.5$).

\begin{figure}[htb]
\centerline{\psfig{file=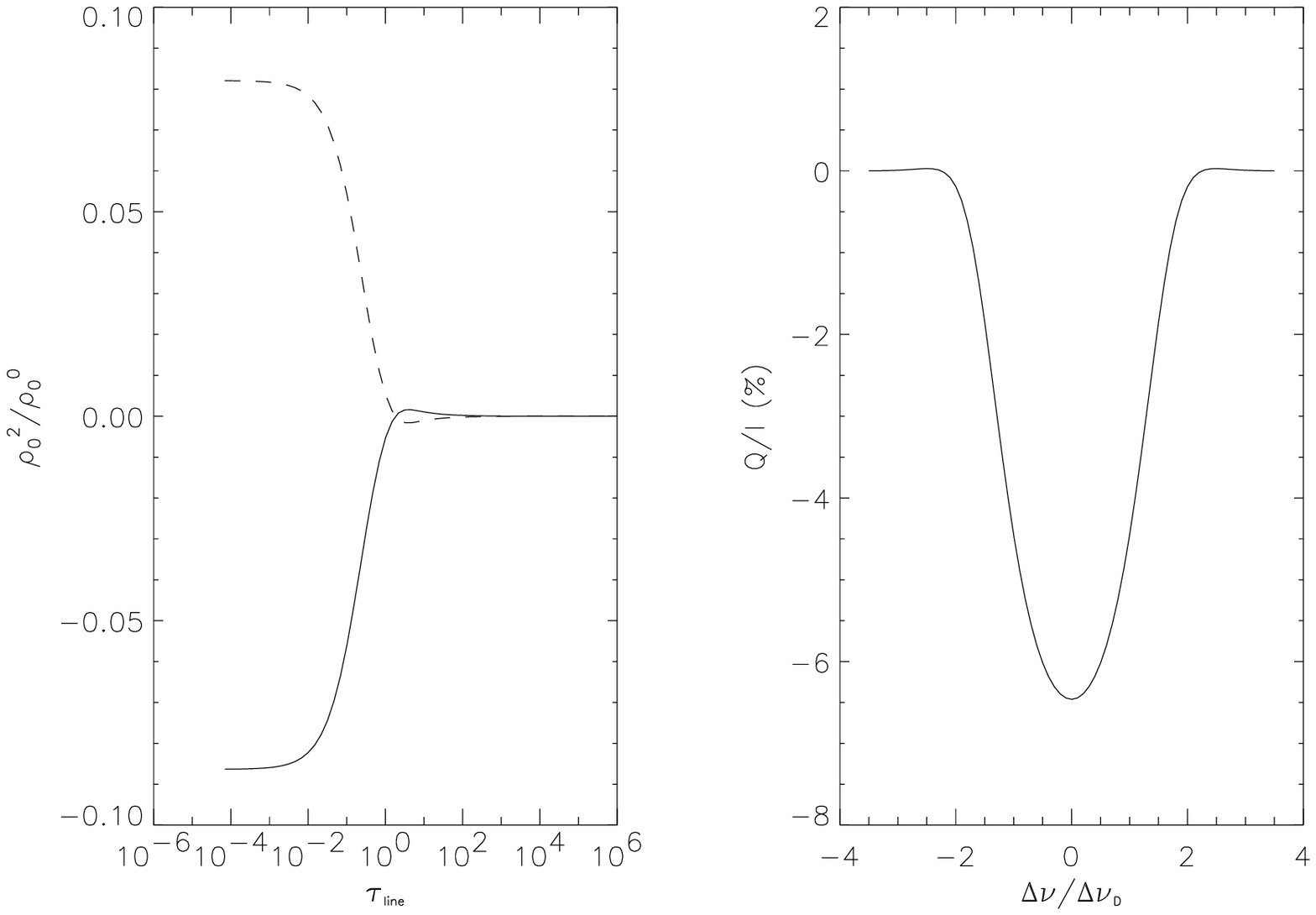,width=12cm,height=7cm}}
\label{fig1}
\caption[fig1]{Predictions of the full theory 
for line transitions in a two-level model atom
with ${\rm J}_l={\rm J}_u=1$. The left panel (Fig. 3$a$) 
shows the variation with line integrated optical depth of 
$\rho^2_0(l)/\rho^0_0(l)$ (dashed line) and of
$\rho^2_0(u)/\rho^0_0(u)$ (solid line). The right panel (Fig. 3$b$) 
shows the emergent fractional linear polarization
at $\mu=0.1$, which is to be compared with the dashed-line of Fig. 1$b$.}
\end{figure}

Figure 3$a$ shows, for the {\it zero}
depolarizing rate case of  
$1\,\rightarrow\,1$ transitions, 
the self-consistent solution for $\rho^2_0(u)/\rho^0_0(u)$ (solid line)
and $\rho^2_0(l)/\rho^0_0(l)$ (dashed line). As with the previous
example, I solved this non-linear and non-local
non-LTE problem of the $2^{nd}$ kind by applying the iterative
methods described in the Appendix. The ensuing
emergent fractional linear polarization at $\mu=0.1$ is 
given in Figure 3$b$. 
We see, again, that the prediction of the correct 
two-level atom theory for line transitions with 
${\rm J}_l={\rm J}_u=1$ is that the emergent linear polarization
profile {\it is as important as that corresponding to triplet lines}.
In conclusion, in the absence of relaxation mechanisms,
three different types of line transitions (which according to the
{\it standard} theory should show up very different degrees
of linear polarization) turn out to lead to similar
amounts of linear polarization when the effect of
lower-level atomic alignment is taken into account.

\section{The Effect of Depolarizing Elastic Collisions}

Elastic collisions have a depolarizing
role, i.e. they reduce the alignment of the atomic levels.
The rate of level depolarization is expected to be comparable
with the ordinary collisional broadening rate. 
Figure 4 shows a rough estimate of the variation 
with height in the VAL-C solar atmospheric model 
(Vernazza, Avrett and Loeser, 1981) of the depolarizing
rates corresponding to the lower and upper levels of the Mg I $b_2$ line,
whose Einstein coefficient for spontaneous emission is $A_{ul}=10^8\,{\rm s}^{-1}$. The approximate formulae for the calculation of
depolarizing rates used here are similar
to those presented by Lamb and Ter Haar (1971), which 
only take into account elastic collisions with
neutral hydrogen atoms and are based
on the short range of the van der Waals interaction.  
For reference purposes the dotted line gives the 
height variation of the temperature in the VAL-C solar model atmosphere.
As seen in the figure the rough estimates of the depolarizing rates 
for the chromospheric Mg $b_2$ line
vary between $10^8\,{\rm }{\rm s}^{-1}$ in the solar photosphere
and $10^3\,{\rm s}^{-1}$ in the upper chromosphere, with a value
of about $10^7\,{\rm s}^{-1}$ at a height of 500 km.

\begin{figure}[htb]
\centerline{\psfig{file=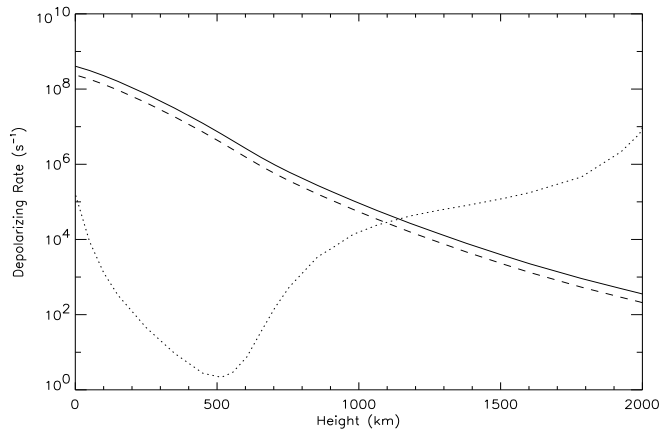,width=9cm,height=6cm}}
\label{fig1}
\caption[fig1]{A rough estimate of the height
variation of the depolarizing rates in the VAL-C
solar model atmosphere (see the dotted line for its temperature profile)
for the lower level (dashed line)
and the upper level (solid line) of the Mg I $b_2$ line, 
which has $A_{ul}=10^{8}\,{\rm s}^{-1}$.}
\end{figure}

\begin{figure}[htb]
\centerline{\psfig{file=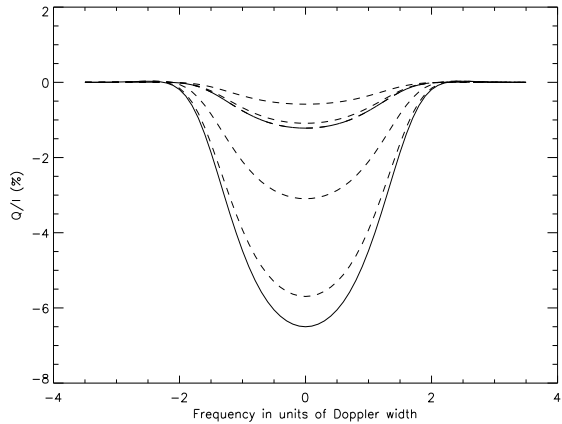,width=8cm,height=6cm}}
\label{fig1}
\caption[fig1]{The emergent fractional linear polarization
for line transitions
with ${\rm J}_l={\rm J}_u=1$, and for various depolarizing rate values
$\delta$ (measured in units of the Einstein $A_{ul}$ coefficient).
The dashed lines show the emergent Q/I profiles for
$\delta=10^{-5}$, $\delta=10^{-4}$,
$\delta=10^{-2}$, $\delta=10^{-1}$
and $\delta=1$, from the largest to the smallest plotted
linear polarization signals, respectively. 
The solid line refers to the zero depolarizing rate case.}
\end{figure}

Figure 5 shows, for the case of line transitions with ${\rm J}_l={\rm J}_u=1$,
the emergent fractional linear polarization at $\mu=0.1$ for the 
$D_u=D_l=0$ depolarizing rate case (solid line), and for increasing values
of $\delta=D/A_{ul}$, with $D=D_u=D_l$ and $A_{ul}$ the Einstein
coefficient for spontaneous emission. As expected, increasing $\delta$-values
produce correspondingly decreasing
emergent fractional linear polarization signals. 
Note that for the case with $\delta=10^{-2}$
(see the long-dashed line)
the resulting Q/I is similar to the result provided by the standard
theory, which neglects lower-level polarization (see Fig. 1$b$). In other
words, for $\delta=10^{-2}$ the alignment of the
ground level has been already destroyed, while the 
amount of upper-level alignment
is still important and very similar to that predicted by the standard
theory. The conclusion is that 
the alignment of the {\it ground} level of {\it optical} 
resonance-line transitions
is very sensitive to depolarizing elastic collisions and that
{\it only if} $\delta\,<\,10^{-5}$ {\it can we expect to
have the maximum} Q/I {\it signal possible}.

\section{A Possible Multilevel Solution to the Na Solar Paradox}

The explanation of the linear polarization pattern around the
Na ${\rm D}_2$ and ${\rm D}_1$ lines proposed by Landi Degl'Innocenti (1998)
implies the {\it absence} of depolarizing effects due to magnetic
fields and/or to elastic collisions. 
As pointed out by Landi Degl'Innocenti (1998) this apparently leads
to a paradox because, on the one hand, his two-level
atom modelling, characterized by an amount of {\it ground}-level atomic polarization larger than that of the upper level,
beautifully explains the polarization profile of the Na doublet.
On the other hand, this 
{\it ground}-level polarization cannot survive in the presence
of turbulent or canopy-like horizontal
fields stronger than 0.01 gauss, which contradicts
previous evidence (see Jones, 1984; Solanki and Steiner, 1990;
Faurobert-Scholl, 1992; Bianda {\it et. al}, 1998; Stenflo {\it et. al}, 1998). 
In principle, the presence of atomic polarization
in the {\it ground} level of Na I
also seems difficult to understand because,
as shown above, even depolarizing 
collisional rates 
with $\delta$-values as small as $\delta=10^{-4}$ 
would significantly reduce the amount of {\it ground}-level 
atomic alignment that is needed
to fit the observations according to Landi Degl'Innocenti's (1998)
modelling.

Can one think of a {\it multilevel} scenario for Na
that might help resolve 
this ``Na Solar Paradox''? The answer is affirmative.
The following multilevel scenario for Na is motivated by the
two-level atom calculations presented in Section 3 and also by
multilevel Na modelling without polarization physics.
In fact, it is known that the two-level atom approximation
does not apply to minority species like Na I, because the difference
in coupling to the continuum reservoir between the two levels
affects the populations strongly (Bruls, Rutten and Shchukina, 1992).
The two-level atom rate Eqs. (6) and (11) can also be invoked to point out
that the amount of atomic alignment
of the upper level of a line transition 
in a {\it multilevel} atomic model is expected to be {\it different} from
that predicted by a two-level atom approach. On the one hand, 
{\it extra} contributions to $\rho^2_0(u)$ 
(with $u$ the upper level of the line transition
being considered) can come from the 
transfer of atomic alignment from lower levels ``$i$'' 
with excitation energy ${\rm E}_i<{\rm E}_u$, including also
the contribution due to the $K=2$ multipole component of the
upward inelastic collisional rate ($C_{i,u}^{(2)}$).
On the other hand, $\rho^2_0(u)$
can be modified via the 
radiative rates that are due to {\it absorption}
to higher levels ``$j$'' with ${\rm E}_j>{\rm E}_u$ and also because,
in a multilevel model atom, 
we can have transfer of alignment
from higher levels ``$j$'' 
to the level ``$u$'', including also contributions that
can arise via the $K=2$ multipole component of the
downward inelastic collisional rate ($C_{j,u}^{(2)}$).

For the Na ${\rm D}_2$ and ${\rm D}_1$ lines their lower level 
$^2 \rm S_{1/2}$ is the ground level
and in the multilevel scenario being suggested here there is no
atomic polarization in this {\it ground} level because it is completely
destroyed by both magnetic fields and elastic collisions. The extra
contributions to the alignment of the 
hyperfine components of the upper levels $^2 \rm P_{3/2}$
and $^2 \rm P_{1/2}$ of the ${\rm D}_2$ and ${\rm D}_1$ lines, 
which is needed to
explain the observations, can then only come from the transfer of alignment
from higher levels and {\it from the relaxation rates due to
absorption processes from these two upper levels to higher ones}.
By having a detailed look at multilevel atomic models
for Na it is easy to see that there are radiative transitions 
from the levels $^2 \rm P_{3/2}$ and $^2 \rm P_{1/2}$
to higher levels in the Grotrian diagram that are 
probably influenced by optical pumping processes.
In doing this complicated
modelling for the Na atom it will be also of interest to 
take into account the transfer of atomic alignment,
among the hyperfine components of the two levels $^2 \rm P_{3/2}$
and $^2 \rm P_{1/2}$, due to the multipole components of 
inelastic collisional rates. 
Since the lifetime of non-ground atomic levels is about two orders
of magnitude smaller than the ground-level lifetime, one finds that
neither low-gauss magnetic fields nor elastic collisions with
rates not larger than $A_{ul}$ can destroy the atomic
polarization of such excited levels. Note that in this multilevel
scenario there is no contribution to the emitted linear polarization
due to dichroism (i.e. to the term $-\eta_{\rm Q}{\rm I}$ of Eq. 8),
simply because we are saying that the Na {\it ground}-level atomic polarization
is practically destroyed by depolarizing mechanisms.

Speculations apart, what it is really important
to emphasize is that a better understanding
of the observed linear polarization pattern around the Na D lines
(and of the second solar spectrum in general!) can only be achieved
after carefully formulating and solving {\it multilevel} non-LTE problems
of the $2^{nd}$ kind, taking fully into account the possibility
of {\it atomic polarization} at {\it all} the atomic levels and
considering the depolarizing role of elastic collisions and magnetic fields.
To this end, it is imperative to develop a general multilevel scattering
line polarization code. For this, we had first to develop 
some suitable iterative methods for the solution of non-LTE problems of the $2^{nd}$ kind that are briefly described in the Appendix.

\section{Observational Search for Lower-level Atomic Polarization
and the Mg Solar Paradox}

In an observational search for lower-level atomic polarization
the first choice of lines to look at
are those suggested in the paper of Trujillo Bueno and Landi Degl'Innocenti (1997), i.e. lines with ${\rm J}_l=1$ and ${\rm J}_u=0$ and ideally corresponding to atoms
devoid of hyperfine structure. These are ``null'' lines because
the standard theory predicts zero linear polarization for them
and the only mechanism capable of producing non-zero line polarization values 
is the possible presence of lower-level atomic polarization.
A useful list with this type of line transition 
can easily be obtained 
by selecting only the unblended lines
from a long list of spectral lines with ${\rm J}_l=1$ and ${\rm J}_u=0$ 
that can be automatically found by computer search in 
the whole spectrum. Jorge S\'anchez Almeida and I followed this strategy and,
in September 1997, we made an observing run
using Semel's technique (Semel, 1994; see also Bianda {\it et. al.}, 1998)
with the Gregory-Coud\'e Telescope operated by G\"ottingen University
at the Spanish Observatorio del Teide (Tenerife, Spain).
Unfortunately, we could not reach any definitive
conclusion because, due to non-linearities 
in the CCD camera used, we failed to reach the good
polarimetric sensitivity that should be achievable with Semel's
technique. A similar observational programe is being pursued
independently by other colleagues using 
the polarimeter ZIMPOL attached to the McMath solar telescope
at Kitt Peak Observatory (Stenflo 1998; private communication).
This is an instrumental set-up that can reach very high polarimetric sensitivity and it is likely that their search will soon lead to some useful results.

Perhaps a good idea for exploring whether lower-level
atomic polarization is at work in the solar {\it photosphere} is that
of performing spectropolarimetric observations in
two lines of the same multiplet, having very similar
line formation properties, but one having ${\rm J}_l=1$ and ${\rm J}_u=0$
(i.e. a ``null'' line)
and the other being a triplet line (i.e. with ${\rm J}_l=0$ and ${\rm J}_u=1$).
An example can be found in the Si I 5701.11 and the Si I 5665.55
lines, respectively. 
It is also of interest to mention that, in our list, there is no ``null'' 
line whose lower level is the ground level. The lower level 
is always an excited level whose lifetime is not much larger than the lifetime 
of the upper level. Although the observation of linear polarization in such 
lines would be an irrefutable proof of the existence of lower-level atomic
polarization, the detection of Hanle depolarization in the same lines would 
imply the existence of magnetic fields of the order of a few gauss. The
situation is different from the case analyzed by Landi Degl'Innocenti (1998)
because, for the sodium D lines, the lower level is the ground level
and a magnetic field of the order of few mgauss is sufficient to destroy
its atomic polarization.

One can also try to obtain some observational hints about the
importance of lower-level atomic polarization for understanding
the second solar spectrum by using different line transitions.
For instance, in this paper we have also paid some attention to lines
with ${\rm J}_l={\rm J}_u=1$. A very interesting line with these total angular momentum
values is the Mg I $b_2$ line at 5172.68 \AA, which belongs to the multiplet
$^{3}{\rm P}^{\rm o}\,-\,^{3}{\rm S}$. In this same multiplet we find
the Mg I $b_1$ line at 5183.60 \AA $\,$ (which is a 
${\rm J}_l=2{\rightarrow}{\rm J}_u=1$ transition) and the Mg I $b_4$ line at 5167.32 \AA
$\,$ (which is a ${\rm J}_l=0{\rightarrow}{\rm J}_u=1$ transition). We point out
that $90\%$ of Mg has zero nuclear spin.

These three lines
share the same upper level ($^3 {\rm S}_1$)
that has ${\rm J}_u=1$. Assume that, at each 
point in the solar atmosphere, we know the exact value of the
fractional alignment coefficient
($\beta_u=\rho^2_0(u)/\rho^0_0(u)$) of this upper level. If one now
calculates at the line centre the emergent fractional linear polarization
(Q/I) assuming that there is no atomic polarization in their
corresponding lower levels one finds 
values proportional to $-0.1$, $+0.5$ and $-1$
for the Mg $b_1$, $b_2$ and $b_4$ lines, respectively, and where the minus
sign indicates that the electric vector is parallel to the nearest
solar limb, while the plus sign refers to the perpendicular direction.
However, this prediction of the standard polarization
transfer theory is not correct because old polarimetric
observations (Stenflo {\it et. al.}, 1983) and recent 
ZIMPOL observations of the second solar spectrum
(Stenflo, 1998; private communication) show that the
emergent fractional linear polarization in the Mg $b_1$, $b_2$ and $b_4$
lines are {\it similar}. It is informative to mention that
all the workshop participants could see the Q/I plot of the
${\rm J}_l=0{\rightarrow}{\rm J}_u=1$ Mg I $b_4$ line stamped on 
our workshop identification cards. 

What emergent Q/I signal would we find if 
in the {\it lower} levels of the Mg $b_1$ and $b_2$ lines 
(which have ${\rm J}_l=2$ and ${\rm J}_l=1$, respectively) we had
an amount of atomic alignment {\it larger} 
than the one of the upper level ?
As discussed above, the presence of atomic alignment
in the lower level of a line transition leads to an extra
contribution to the linear polarization originating at each atmospheric point
that comes from the term $-\eta_{\rm Q}{\rm I}$
of Eq. (8), with $\eta_{\rm Q}\,{\sim}\,{\cal Z}\rho^2_0(l)$ (see Eq. 10).
This contribution
adds to the usual one ($\epsilon_{\rm Q}$) that is proportional
to $\omega\,\rho^2_0(u)$ (see Eq. 4). It is not difficult to show that
the following expression holds at $\tau_\nu\approx1$
for the purpose of estimating
the {\it emergent} fractional linear polarization:

\begin{equation}
{{{\eta}_{\rm Q}{\rm I}}\over{{\epsilon}_{\rm Q}}}\,{\approx}\,
{{\cal Z}\over{\omega}}\,{{\beta_l}\over{\beta_u}},
\end{equation}
where $\beta_l=\rho^2_0(l)/\rho^0_0(l)$ and $\beta_u=\rho^2_0(u)/\rho^0_0(u)$.
Using this expression and Eqs. (3), (4) and (10)
we find that, close to the limb,
the {\it emergent} fractional linear polarization
at the line centre is approximately given by

\begin{equation}
{{\rm Q}\over{\rm I}}\,{\approx}\,-{\omega}{\beta_u}\,+\,{\cal Z}{\beta_l},
\end{equation}
where the second contribution is that due to {\it dichroism}.
Taking into account, 
from Table I of Landi Degl'Innocenti (1984),
that ${\cal Z}={\omega}=-0.5$ for the $1{\rightarrow}1$ $b_2$ line and that
${\cal Z}\,{\approx}\,{0.6}$ and $\omega\,=\,0.1$ for the $2{\rightarrow}1$ $b_1$ line, 
and assuming that ${\mid}{\beta_l}{\mid}>{\mid}{\beta_u}{\mid}$
for the Mg $b_1$ and $b_2$ lines,
one arrives at the conclusion that the
presence of a sizable amount of lower-level atomic alignment
would imply that the
emergent fractional linear polarization in the Mg $b_1$, $b_2$ and $b_4$
lines would be {\it similar}, as the above-mentioned observations show.

Finally, it is necessary to note that the lower levels of the Mg $b$ lines
are metastable. The $^3 {\rm P}_1$ level is indeed connected to the
ground level $^1 \rm S_0$ by a forbidden transition at 4571 \AA, but its 
lifetime for spontaneous de-excitations is of the order of 
$5 \times 10^{-3} {\rm s}$, whereas the lifetime of the other two levels 
($^3 {\rm P}_0$ and $^3 {\rm P}_2$) is even larger. Thus, if levels 
$^3 {\rm P}_2$ and $^3 {\rm P}_1$ 
(the lower levels of the $b_1$ and $b_2$
lines, respectively) turn out to have an amount of atomic 
polarization that is {\it larger} (in absolute value)
than that of the upper level, 
we would end up with a conclusion similar to that found by
Landi Degl'Innocenti (1998) using the Na D lines argument, i.e. that
magnetic fields stronger than about a few mgauss, either in the form of
{\it volume-filling}
turbulent fields or in the form of canopy-like, horizontal fields,
cannot exist in the solar chromosphere. 
Although the line formation regions of the
Mg $b$ lines are much {\it higher} than those of the Mg 4571 \AA $\,$ line,
I think that careful spectropolarimetric limb observations
in the Mg 4571 ${\rm J}_l=0{\rightarrow}{\rm J}_u=1$ 
line should be carried out urgently.
This may tell us whether the {\it lower} level of the
Mg $b_2$ line is polarized at the atmospheric heights of formation
of the Mg 4571 \AA $\,$ line ($\sim 400$ km). One should keep in 
mind, however, that this line is formed much closer to LTE 
than other lines of similar strength because it is an
optically forbidden intercombination line.
In any case, a new paradox seems
to exist: the ``Mg solar paradox''.

\section{Concluding Remarks}

There is a crucial difference between the Na and Mg ``solar paradoxes''.
For Na there was still the chance that the multilevel scenario 
outlined above might help to solve it. However, there is
no similar multilevel solution for the ``Mg solar paradox'',
which leads to the conclusion that there must indeed exist {\it ground}-level
and {\it metastable}-level atomic polarization in the solar chromosphere. 
The three Mg $b$ lines share the same upper level and what cannot be understood by means of the standard transfer theory (which neglects lower-level
polarization) is that the observed linear polarization in these three 
Mg $b$ lines turns out to be similar. Thus, even if one chooses
multilevel atomic models for Mg to calculate the 
self-consistent values of the density-matrix elements, 
and then solves the Stokes transfer equations to get
the emergent fractional linear polarization, but 
neglecting the dichroism contribution that comes from the
atomic alignment of the lower {\it metastable}
levels, one would find again that 
the ensuing prediction is wrong. The only way
I see for increasing the emitted polarization in the Mg $b_1$ and
$b_2$ lines, so as to bring it to the same level of that
corresponding to the Mg $b_4$ line (that has ${\rm J}_l=0$), is via
the {\it dichroism} contribution (i.e. the term $-\eta_{\rm Q}{\rm I}$ of
Eq. 8). As discussed in Section 3, this {\it dichroism}
can only arise if the magnetic sublevels of the
${\rm J}_l=2$ and ${\rm J}_l=1$ {\it metastable} lower-levels of the $b_1$ and $b_2$ lines
are {\it unequally populated}.
The same explanation can be given
for other groups of lines belonging to other atoms,
and arising from a similar multiplet ($^3{\rm P}^{\rm o}\,-\,^3{\rm S}$),
with the ensuing appearance of extra ``paradoxes'' for other chemical
elements.
 
As we have seen, the anisotropic illumination of the atoms
in a stellar atmosphere can lead to large population imbalances
among the lower-level sublevels of many spectral lines. 
The modelling of the second solar spectrum
requires the reliable calculation of
the atomic polarization of the lower and upper levels
corresponding to the line transitions of interest.
To this end, it is crucial to be able to consider
{\it multilevel} atomic models. 
This goal can presently be achieved
by formulating the problems of interest
within the framework of the density matrix
polarization transfer theory 
(see Landi Degl'Innocenti, 1983) and
by numerically solving the ensuing non-linear
and non-local equations with the iterative methods
presented in the Appendix.
Only when we know the {\it self-consistent}
values of the alignment coefficients of
the Na and Mg atomic levels in several solar atmospheric models
shall we be able to figure out a possible solution
to such ``solar paradoxes''. 

Obviously, we are facing
a complex problem here, from the observational, theoretical
and modelling viewpoints. But it is a highly interesting one, 
not only because of the fascinating physics
that it involves, but mainly because in trying to clarify it we may learn
something new about the sun.

%%%%%%%%%%%%%%%%%%%

\acknowledgements{\footnotesize 
I am very grateful to Prof. E. Landi Degl'Innocenti 
for several clarifying discussions
and for his careful reading of my paper. Thanks are also due to
Dr. J. S\'anchez Almeida for his help during
the observational search for lower-level atomic polarization
and to Prof. J. Stenflo for kindly sending me Q/I plots of his
ZIMPOL observations of the Mg $b$ lines prior to publication.
Finally, I thank the referee of this paper
(Dr. V. Bommier) for some useful suggestions.
Partial support by the Spanish DGES through project PB 95-0028
is gratefully acknowledged.
This work is part of a continuing
EC-TMR European Solar Magnetometry Network.
}

\vspace{1cm}

\centerline {\bf APPENDIX}
\vspace{0.5cm}

\centerline {\bf Iterative methods for the non-LTE problem
of the $2^{nd}$ kind}
\vspace{0.2cm}

In general, both for two-level and multilevel atomic models, with or without
hyperfine structure included, but taking fully into account {\it atomic
polarization} in {\it all} the atomic levels,
the above-mentioned {\it non-linear} 
and {\it non-local} system of equations can be symbolically represented as

\begin{equation}
\bf A\,x\,=\,b,
\end{equation}
with $\bf b$ a known vector and $\bf x$ the {\it unknown} vector formed
by the density matrix elements at {\it all} the spatial grid-points, and
where $\bf A$ is an operator which depends on
collisional rates (both inelastic and elastic)
and on radiation field tensors
that are given by weighted frequency and angular averages of the Stokes parameters. In these SE equations one finds {\it non-linear} terms of the form
${\bar{J}}^K_Q\,{\rho}^k_q$ (see Eqs. 6 and 11).
They are non-linear because
the radiation field tensors ${\bar{J}}^K_Q$ depend implicitly
on the density matrix elements ${\rho}^k_q$ via the RT equations.
This non-linearity means that
the operator $\bf A$ depends implicitly on the unknown $\bf x$.
The solution of non-linear problems necessarily 
requires the application
of iterative methods. Here, at each iterative step, one has to 
manage to set up and solve a suitable
linear system of equations whose solution leads to approximate
corrections to the unknowns (see Socas Navarro and Trujillo Bueno, 1997).
To this end, it is very important that the approximations one introduces
for achieving linearity at each iterative step adequately treat the
{\it coupling} between transitions and the {\it non-locality} of the problem.

It is also very important to point out that
iterative schemes that require the construction and inversion of large matrices
are useless. Thus, my first step 
towards the modelling of the second solar spectrum has been 
the development of two effective iterative methods that are indeed
capable of solving non-linear polarization transfer 
multilevel problems of the $2^{nd}$ kind without having to build and
invert large matrices at each iterative step. I have called these
methods the $\bf DALI$ and $\bf DEGAS$ iterative schemes. $\bf DALI$,
besides the name of the famous Spanish painter,
is an acronym for Density-matrix ALI method (which is based on Jacobi iteration), while $\bf DEGAS$, besides the name of the fine French painter,
refers here to my Density-matrix Gauss-Seidel iterative scheme.

As pointed out above the non-linear terms appearing in the SE equations
are of the form ${\bar{J}}^K_Q\,{\rho}^k_q$. 
A detailed presentation of these
iterative methods will be published elsewhere.
Here I simply give a
``numerical recipe'' for implementing 
the DALI and DEGAS methods. It consists in making the following
changes for achieving linearity 
in the SE equations at each iterative step:

\vskip 0.5truecm

If K$\ne$0,

\begin{equation}
{{{\bar{J}}^K_Q}}\,\,{\rho}^k_q\,\,\rightarrow\,\,
{{{\bar{J}}^K_Q}}{^{old}}\,\,{{\rho}^k_q}^{{new}}
\end{equation}

\vskip 0.5truecm

If K=0,

\begin{equation}
{{\bar J}^0_0}\,\,{\rho}^k_q\,\,\rightarrow\,\,
{{\bar J}^0_0}{^{*}}\,\,{{\rho}^k_q}^{new}\,+\,
{{{\Lambda}^0_0}(i,i)}\,[\,{{\rho}^k_q}^{{old}}
{{\rho}^0_0}^{{new}}({u})\,\,-\,\,
{{\rho}^0_0}^{{old}}({u})\,{{\rho}^k_q}^{{new}}\,],
\end{equation}  
where ${{\Lambda}^0_0}(i,i)$ (``$i$'' being the spatial grid-point
under consideration) is the {\it diagonal} element
of a $\Lambda-$ operator that arises in the definition of 
${\bar{J}}^0_0$, and where
``$old$'' is meant to take the value of the previous iterative step,
while ``$\rho^k_q{^{new}}$'' simply indicates the 
density-matrix elements that are to be obtained at the current iterative step
by simply solving the resulting linear system of equations.
In the DALI method we take
${{{\bar{J}}^0_0}}{^{\,{*}}}\,=\,{{{\bar{J}}^0_0}}{^{{old}}}$,
while for DEGAS 
${{{\bar{J}}^0_0}}{^{\,{*}}}\,=\,{{{\bar{J}}^0_0}}{^{old\,{\rm and}\,new}}$,
with these two quantities (and also ${{\Lambda}^0_0}(i,i)$)
calculated, at each iterative step, as explained
by Trujillo Bueno and Fabiani Bendicho (1995) (see also Trujillo Bueno and
Manso Sainz, 1999). 

\begin{figure}[htb]
\centerline{\psfig{file=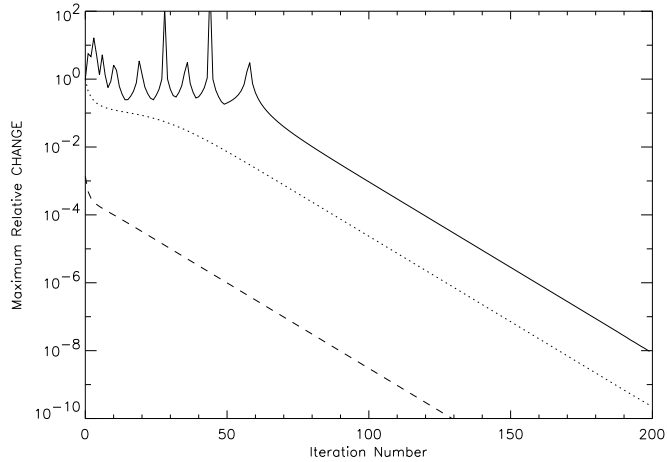,width=9cm,height=6.5cm}}
\label{fig1}
\caption[fig1]{The convergence rate of the DALI iterative method
applied for numerically solving 
the ${\rm J}_l={\rm J}_u=1$ scattering line polarization problem
taking into account atomic alignment in both levels. The solid
line gives the variation with the iterative step of the
maximum relative change in $\rho^2_0(u)$ and $\rho^2_0(l)$,
the dotted line in $\rho^0_0(u)$ and the dashed line in $\rho^0_0(l)$.}
\end{figure}

Figure 6 shows an example of the convergence rate
of the DALI method for a calculation initialized using the LTE
density-matrix $\rho$-values. With the DEGAS multilevel iterative scheme the
number of iterations (and the computing time)
required to achieve convergence is substantially smaller.

For the solution of scattering line polarization
problems using realistic multilevel atoms it is better to initialize
the calculation using the $\rho^0_0$ self-consistent values
corresponding to the {\it unpolarized} case. 
Since these $\rho^0_0$-values can be directly obtained
from the atomic level populations such an initialization can be found
easily by using any of the fast RT multilevel codes that are now available
(see, e.g., Socas Navarro and Trujillo Bueno, 1997; Fabiani Bendicho
and Trujillo Bueno, 1999). With this initialization given, it is possible
to obtain the self-consistent solution of multilevel scattering line polarization problems formulated with the density matrix theory applying
{\it one} of the three following methods:
$\Lambda-$iteration, DALI iteration
or DEGAS iteration, with increasing
improvements in the resulting convergence rate.


\begin{thebibliography}{}
\bibitem[]{}

\bibitem[]{}
Bianda, M., Solanki, S.K., Stenflo, J.O. (1998), {\it Astron. Astrophys.},
{\bf 331}, 760
\bibitem[]{}
Bommier, V. (1997a), {\it Astron. Astrophys.}, {\bf 328}, 706
\bibitem[]{}
Bommier, V. (1997b), {\it Astron. Astrophys.}, {\bf 328}, 726
\bibitem[]{}
Bommier, V. Sahal-Br\'echot (1978), {\it Astron. Astrophys.}, {\bf 69}, 57
\bibitem[]{}
Bruls, J., Rutten, R.J., Shchukina, N.G. (1992), {\it Astron. Astrophys.}, {\bf 265}, 237
\bibitem[]{}
Fabiani Bendicho, P., Trujillo Bueno, J. (1999), in {\it Solar Polarization}, edited by K.N. Nagendra \& J.O. Stenflo. Kluwer Academic Publishers, 1999. (Astrophysics and Space Science Library ; V. 243), p. 219-230
\bibitem[]{}
Faurobert-Scholl, M. (1992), {\it Astron. Astrophys.},{\bf 258}, 521
\bibitem[]{}
Happer, W. (1972), {\it Rev. Mod. Phys.}, {\bf 44}, 169
\bibitem[]{}
Jones, H. P. (1984), in ``Chromospheric Diagnostics and Modelling'',
B. Lites (ed.), National Solar Observatory
\bibitem[]{}
Lamb, F. K., Ter Haar, D. (1971), {\it Physics Reports}, {\bf 2}, No. 4, 253
\bibitem[]{}
Landi Degl'Innocenti, E. (1983), {\it Solar Phys.}, {\bf 85}, 3
\bibitem[]{}
Landi Degl'Innocenti, E. (1984), {\it Solar Phys.}, {\bf 91}, 1
\bibitem[]{}
Landi Degl'Innocenti, E. (1987), in {\it Numerical Radiative Transfer},
W. Kalkofen (ed.), Cambridge University Press, 265
\bibitem[]{}
Landi Degl'Innocenti, E. (1998), {\it Nature}, {\bf 392}, 256
\bibitem[]{}
Landi Degl'Innocenti, E., Bommier, V., Sahal-Br\'echot, S. (1990), {\it Astron. Astrophys.}, {\bf 235}, 459
\bibitem[]{}
Landi Degl'Innocenti, E., Landi Degl'Innocenti, M., Landolfi, M. (1997),
in {\it Science with Themis}, N. Mein and Sahal-Br\'echot (eds.),
Paris Obs. Publ., p. 59
\bibitem[]{}
Landolfi, M., Landi Degl'Innocenti, E. (1986), 
{\it Astron. Astrophys.},{\bf 167}, 200
\bibitem[]{}
Semel, M. (1994), in {\it Solar Surface Magnetism}, R.J. Rutten
and C.J. Schrijver (eds.), Kluwer, p. 509
\bibitem[]{}
Socas Navarro, H., Trujillo Bueno, J. (1997), {\it Astrophysical
Journal}, {\bf 490}, 383
\bibitem[]{}
Solanki, S., Steiner, O. (1990), {\it Astron. Astrophys.}, {\bf 234}, 519
\bibitem[]{}
Stenflo, J.O. (1994) {\it Solar Magnetic Fields. Polarized Radiation
  Diagnostics}. Kluwer Academic Publishers, Dordrecht.
\bibitem[]{}
Stenflo, J.O. (1997), {\it Astron. Astrophys.}, {\bf 324}, 344
\bibitem[]{}
Stenflo, J.O., Twerenbold, D., Harvey, J. W., Brault, W. (1983),
{\it Astron. Astrophys. Suppl. Ser.}, {\bf 54}, 505
\bibitem[]{}
Stenflo, J.O., Keller, C.U. (1997), {\it Astron. Astrophys.}, {\bf 321}, 927
\bibitem[]{}
Stenflo, J.O., Keller, C.U., Gandorfer, A. (1998), {\it Astron. Astrophys.},
{\bf 329}, 319
\bibitem[]{}
Trujillo Bueno, J. and Fabiani Bendicho P. (1995), 
{\it Astrophysical Journal}, {\bf 455}, 646
\bibitem[]{}
Trujillo Bueno, J. and Landi Degl'Innocenti, E. (1997), {\it Astrophysical
Journal Letters}, {\bf 482}, 183
\bibitem[]{}
Trujillo Bueno, J., Manso Sainz, R. (1999), {\it Astrophys. Journal}, {\bf 516}, 436
\bibitem[]{}
Vernazza, J., Avrett, E., Loeser, R. (1981), {\it Astrophys. J. Suppl. Series},
{\bf 45}, 635

\end{thebibliography}
\end{document}